\documentclass[prd,preprint,showpacs,preprintnumbers,amsmath,amssymb,eqsecnum,nofootinbib,groupedaddress,superscriptaddress,pdflatex]{revtex4-1}
\usepackage{amsfonts,mathrsfs}
\usepackage[dvipdfmx]{hyperref,graphicx,color}

\begin{document}
\preprint{RUP-23-10}
%
\title{
Revisiting compaction functions for primordial black hole formation}
\author{Tomohiro Harada}
\email{harada@rikkyo.ac.jp}
\affiliation{Department of Physics, Rikkyo University, Toshima,
Tokyo 171-8501, Japan}
\author{Chul-Moon Yoo}
\email{yoo.chulmoon.k6@f.mail.nagoya-u.ac.jp}
\affiliation{Graduate School of Science,
Nagoya University, Nagoya 464-8602, Japan}
\author{Yasutaka Koga}
\email{y.koga.tj@gmail.com}
\affiliation{Graduate School of Science,
Nagoya University, Nagoya 464-8602, Japan}
\begin{abstract}
Shibata and Sasaki~[Phys. Rev. D \textbf{60}, 084002 (1999)] 
introduced the so-called compaction function.
Since then, it has been empirically established that 
the maximum value of this function (or its volume-averaged counterpart)
in the long-wavelength solutions gives a very robust threshold of 
primordial black hole formation. In this paper, 
we show that in spite of initial intention, 
the Shibata-Sasaki compaction function cannot be 
interpreted as the ratio of the mass excess to the areal radius 
in the constant-mean-curvature slice of their choice 
but coincides with that in the {\it comoving} slice 
up to a constant factor depending on the equation of state. 
We also discuss the gauge-(in)dependence of 
the legitimate compaction function, i.e., the ratio of the mass excess to the areal radius, 
in the long-wavelength solutions.
\end{abstract}

\date{\today}

\maketitle

\tableofcontents

\newpage

\section{Introduction}

Primordial black holes are black holes that may have formed in the early Universe not as 
the products of the stellar evolution.
These black holes may have remained until now if their masses are 
greater than $10^{15}$ g and currently may play roles 
as dark matter and sources of gravitational sources, while those of masses smaller than that 
may affect the thermal history of the Universe. 
Therefore, primordial black holes can be observed in principle and 
even no detection of them is of significant interest.
See Ref.~\cite{Carr:2020gox} and references therein for a recent review on observational constraints
on primordial black holes.

In modern cosmology, primordial black holes play important roles because
they can carry the information of the early phase of the Universe.
Primordial black holes may have formed in various scenarios but the formation from primordial fluctuations has been most extensively studied so far. Quantum fluctuations generated in an inflationary era of the Universe and stretched to super-horizon scales may have potential to produce primordial black holes.
In the decelerated-expansion era following the inflationary era, 
the fluctuations enter the horizon scale and collapse to black holes. 

In this regime, only high peaks can collapse to black holes if the matter has relativistic pressure. To quantitatively discuss the formation of primordial 
black holes, we need to know under what conditions primordial black holes
can form. In the early studies, the threshold was calculated on the amplitude 
of density perturbation averaged over a horizon patch at its 
horizon entry by Carr~\cite{Carr:1975qj} and later refined in Ref.\cite{Harada:2013epa}
based on analytical argument.
On the other hand, the threshold of this value has been systematically 
obtained through numerical simulations of the Einstein equation with relativistic fluids (e.g.~\cite{Niemeyer:1999ak,Musco:2004ak,Musco:2012au}).

Another quantity, the maximum value of the so-called compaction function
has been proposed as 
the one giving a robust threshold of primordial black hole formation by Shibata and 
Sasaki~\cite{Shibata:1999zs} through general relativistic numerical simulations.
Here we call this function the Shibata-Sasaki compaction function.
Subsequently, this has been shown to be closely related to the density perturbation averaged over a horizon patch in Harada, Yoo, Nakama and Koga~\cite{Harada:2015yda}.
The Shibata-Sasaki compaction function is still advantageous to the averaged density perturbation because of its much simpler definition.
Recently, the volume-averaged Shibata-Sasaki compaction function has been 
proposed as a further robust quantity to give a threshold~\cite{Escriva:2019phb}.
See Ref.~\cite{Escriva:2021aeh} and references therein for a recent review on general relativistic numerical simulation of 
spherically symmetric formation of primordial black holes.

In this paper, we show that the Shibata-Sasaki compaction function introduced in 
Ref.~\cite{Shibata:1999zs}
is not what they intended. Thus, this paper corrects the misunderstanding in
not only Ref.~\cite{Shibata:1999zs} but also Ref.~\cite{Harada:2015yda} and possibly several other papers. 
We should however stress that this does not affect the robustness of the Shibata-Sasaki compaction function. Simultaneously, we clarify 
the relationship between the Shibata-Sasaki compaction function and 
other `legitimate' compaction functions and identify the uniqueness of the former which makes it so robust. This paper is organised as follows. Section~\ref{sec:preliminaries} gives preliminaries for later discussion. In Sec.~\ref{sec:SS}, we introduce the Shibata-Sasaki compaction function. 
In Sec.~\ref{sec:SS_revisited}, we revisit this function and clarify what it is. 
In Sec.~\ref{sec:SS_gauge}, we analytically 
reconstruct the full set of long-wavelength solutions in the Shibata-Sasaki gauge choice.
Section~\ref{sec:summary} 
is devoted to summary. We use the geometrised units in which $G=c=1$ and 
sign conventions in Wald~\cite{Wald:1984rg}.

\section{Preliminaries \label{sec:preliminaries}}

The line element in the flat Friedmann-Lema\^{i}tre-Robertson-Walker (FLRW) 
spacetime can be written in the following form:
\begin{equation}
 ds^{2}=-dt^{2}+a^{2}(t)(dx^{2}+dy^{2}+dz^{2}).
\end{equation}
The spacetime generally begins with a big-bang singularity at $t=0$, where $a=0$.

There is a framework of solutions, the so-called long-wavelength 
solutions~\cite{Shibata:1999zs,Lyth:2004gb}. They 
have the following asymptotic form: 
\begin{equation}
 ds^{2}\approx -dt^{2}+a^{2}(t)\Psi^{4} (x,y,z)(dx^{2}+dy^{2}+dz^{2}) 
\end{equation}
as $t\to 0$ with $\Psi-1=O(1)$, which reduces to the form 
\begin{equation}
ds^{2}\approx -dt^{2}+a^{2}(t)\Psi^{4} (r)(dr^{2}+r^{2}d\Omega^{2})
\end{equation}
in spherical symmetry, where $d\Omega^{2}:=d\theta^{2}+\sin^{2}\theta d\phi^{2}$.
To obtain these solutions, we introduce a smoothing length $L=a/k$.
Then, we assume that the Hubble length $\ell_{H}:=H^{-1}$, where $H=\dot{a}/a$, is much smaller than the smoothing length $L$, so that
\begin{eqnarray}
 \epsilon(t):=\frac{\ell_{H}}{L}=\frac{k}{aH} 
\end{eqnarray}
is much smaller than unity. 
Then, we assume 
\begin{equation}
 \frac{\partial_{i}\ln \Psi}{aH}=O(\epsilon).
\end{equation}
Thus, the field equations can be expanded in terms of $\epsilon$. 
This is called the gradient expansion. 
If $a(t)$ is given by $a=a_{0}t^{\alpha}$, then $\epsilon= [k/(\alpha a_{0})] t^{1-\alpha}$. Thus, in the decelerated expansion, where $0<\alpha<1$, $\epsilon\to 0$ as $t\to 0$, i.e., in the big-bang limit, while in the accelerated expansion, where $\alpha>1$, $\epsilon\to 0$ is realised as $t \to \infty$, i.e., in the late-time limit. 
The long-wavelength solutions have been applied for primordial black hole formation
by Shibata and Sasaki~\cite{Shibata:1999zs}.

In spherical symmetry, there is another framework of solutions, 
in which the line element takes the following form:
\begin{eqnarray}
 ds^{2}\approx -dt^{2}+a^{2}(t)\left[\frac{d\varrho^{2}}{1-K(\varrho)\varrho^{2}}+\varrho^{2}d\Omega^{2}\right]
\label{eq:AGHS} 
\end{eqnarray}
as $t\to 0$.
These solutions are called asympototically quasi-homogeneous 
solutions and have been applied as initial conditions for primordial black hole 
formation in Polnarev and Musco~\cite{Polnarev:2006aa}.
The equivalence between the long-wavelength solutions 
and the asymptotically quasi-homogeneous solutions 
has been shown in Ref.~\cite{Harada:2015yda}.

Although the proposal of primordial black holes goes back to 
Hawking~\cite{Hawking:1971ei} prior to the development of inflationary cosmology since late seventies, the formation scenario from fluctuations generated in the 
inflationary era of the Universe has been extensively discussed. In this context, in the inflationary era, 
quantum fluctuations are stretched to super-horizon scales and become classical perturbations. After the inflation ends, the evolution of the fluctuations is described by the long-wavelength solutions in the decelerated Universe. Finally, the scales of the perturbations get smaller than the horizon scale, which is called horizon entry, and then the long-wavelength scheme breaks down.

For later convenience, we discuss the 
cosmological conformal $3+1$ decomposition of the spacetime
\begin{equation}
ds^{2}=-\alpha^{2}dt^{2}+ \psi^{4}a^{2}(t)\tilde{\gamma}_{ij}(dx^{i}+\beta^{i}dt)(dx^{j}+\beta^{j}dt).
\end{equation}
We impose the condition $\tilde{\gamma}=\eta$, where 
$\tilde{\gamma}$ and $\eta$ are the determinants of $\tilde{\gamma}_{ij}$ 
and a static flat 3-metric $\eta_{ij}$, respectively.
The flat FLRW spacetime corresponds to $\alpha=1$, $\beta^{i}=0$, 
$\psi=1$ and $\tilde{\gamma}_{ij}=\eta_{ij}$.

We make the following additional assumptions. First, we assume that the matter field is 
described by a single perfect fluid, so that 
$T^{\mu\nu}=(\rho+p)u^{\mu}u^{\nu}+pg^{\mu\nu}$ with 
the equation of state $p=(\Gamma-1)\rho$.
We write $\psi(t,{\bf x})=\Psi({\bf x})[1+\xi(t,{\bf x})]$, $\alpha(t,{\bf x})=1+\chi(t,{\bf x})$ 
and $\tilde{\gamma}_{ij}(t,{\bf x})= \eta_{ij}({\bf x})+h_{ij}(t,{\bf x})$. Then, the 
$\epsilon$ expansion inferred from the Einstein equation and the matter 
equation of motion imply the following~\cite{Shibata:1999zs, Lyth:2004gb, Harada:2015yda}. 
For the metric functions, we have $\Psi({\bf x})=O(\epsilon^{0})$, $\xi(t,{\bf x})=O(\epsilon^{2})$,  $\beta^{i}(t,{\bf x})=O(\epsilon)$,
$\chi(t,{\bf x})=O(\epsilon^{2})$ and $h_{ij}=O(\epsilon^{2})$.
For the matter functions, we have 
$\delta(t,{\bf x}):=({\rho-\rho_{b}})/{\rho_{b}}=O(\epsilon^{2})$ and 
$v^{i}(t,{\bf x}):={u^{i}}/{u^{t}}=O(\epsilon)$, where $\rho_{b}=\rho_{b}(t)$ is the 
energy density of the background FLRW solution.
The mean curvature $K$ on the constant $t$ slice follows
$K(t,{\bf x})=K_{b}[1+ \kappa(t,{\bf x})]$, where $K_{b}=K_{b}(t)=-3H$ is the background value and $\kappa(t,{\bf x})=O(\epsilon^{2})$.
From the Einstein equation and the equation of motion, 
we can obtain nonlinearly perturbed solutions order by order in $\epsilon$. 

However, there is a gauge issue in the cosmological long-wavelength solutions.
This comes from the freedom in choosing the lapse function $\alpha$ and the shift vector $\beta^{i}$.  As for $\alpha$, which prescribes the choice of time slicing, we can have the constant-mean-curvature (CMC) slice $\kappa=0$, 
the uniform-density slice $\delta=0$, 
the comoving slice $u_{i}=0$,
the geodesic slice $\chi=0$ and so on.
As for $\beta^{i}$, which prescribes the threading of the constant $x^{i}$ curves, we can have 
the comoving thread $v^{i}=0$, the normal coordinates $\beta^{i}=0$, 
the conformally flat coordinates $h_{ij}=0$ and so on.

The Einstein equations for the long-wavelength solutions
are written down order by order 
in Eqs. (4.12)--(4.26) and 
(4.27)--(4.79) in Ref.~\cite{Harada:2015yda}, respectively,
to the next-to-leading order.
We do not repeat them here but will quote the equations when needed.
In the CMC slice, the solutions for the density perturbation $\delta$ and the covariant component of the 4-velocity $u_{j}$ are given in Eqs.~(4.37), (4.45) and (4.49) in Ref.~\cite{Harada:2015yda} as
\begin{eqnarray}
\delta_{\rm CMC}\approx  f\left(\frac{1}{aH}\right)^{2}, ~~
u_{{\rm CMC} j}\approx \frac{2}{3\Gamma(3\Gamma+2)H}\delta_{{\rm CMC},j},
\label{eq:LWL_sol_CMC}
\end{eqnarray}
where 
\begin{equation}
f: = -\frac{4}{3}\frac{\bar{\Delta}\Psi}{\Psi^{5}} 
\label{eq:f} 
\end{equation}
with $\bar{\Delta}$ being the flat Laplacian.
In the comoving slice, the solutions for $\delta$ and $u_{j}$ are given in Eqs.~(4.68)
and (4.72) in Ref.~\cite{Harada:2015yda} as
\begin{eqnarray}
\delta_{\rm com}\approx \frac{3\Gamma}{3\Gamma+2}f\left(\frac{1}{aH}\right)^{2},~~
u_{{\rm com} j}=0.
\end{eqnarray}
To this order, the choice of the threading condition does not affect the above 
functions of the long-wavelength solutions~\cite{Harada:2015yda}. 
The same functions $\delta$ and $u_{j}$ for 
the geodesic slice and uniform-density slice are presented in Appendix~\ref{sec:G_UD_slices}.

To introduce the compaction function, we should first introduce a quasi-local mass. In spherical symmetry, it is well-known that the 
Kodama mass~\cite{Kodama:1979vn} or the Misner-Sharp mass~\cite{Misner:1964je,Hayward:1994bu} has
physically reasonable properties as a quasi-local mass. In general spherically symmetric spacetime,
 the line element is given by  
\begin{equation}
 ds^{2}=g_{AB}(x^{C})dx^{A}dx^{B}+R^{2}(x^{C})d\Omega^{2},   
\end{equation}
where $A$, $B$ and $C$ run over $0$ and $1$. The Kodama vector $K^{\mu}$, the Kodama current $S^{\mu}$ and the Kodama mass $M_{K}$ are defined as
$K^{\mu}:=-\epsilon^{AB}\partial_{B}R(\partial/\partial x^{A})^{\mu}$, $S^{\mu}:=-T^{\mu}_{~\nu}K^{\nu}$ and  
$M_{K}:=-\int_{\Sigma}S^{\mu}d\Sigma_{\mu}$, respectively, where 
$\epsilon_{AB}$ is the Levi-Civita tensor associated with the 2-dimensional metric $g_{AB}$ 
and $\Sigma$ is the 3-volume on a spherically symmetric spacelike-hypersurface 
with a regular centre
and a boundary $\partial \Sigma$ characterised by $x^{A}$.
On the other hand, we can define the Misner-Sharp mass 
as $M_{\rm MS}:=R(1-D_{A}R D^{A}R)/2$ as a function of $x^{A}$, 
where $D_{A}$ denotes the covariant derivative associated with $g_{AB}$.
We can prove $ M_{\rm MS}=M_{K}$ through 
the Einstein equation by assuming a regular centre. 
See, e.g., Refs.~\cite{Hayward:1994bu,Sato:2022yto,Yoo:2022mzl} for a proof.
Because of this equivalence, 
we simply write $M$ for the Kodama mass and the Misner-Sharp mass.
Thus, the mass $M$ admits not only the expression in terms of the integral of the matter stress-energy tensor over the 3-volume $\Sigma$
\begin{equation}
 M=-\int_{\Sigma}S^{\mu}d\Sigma_{\mu}
\label{eq:M_Kodama}
\end{equation}
but also the expression in terms of the metric functions 
at the boundary $\partial \Sigma$
\begin{equation}
 M=\left.\frac{1}{2}R(1-D_{A}R D^{A}R)\right|_{x^{A}}.
\label{eq:M_MS}
\end{equation}

\section{Shibata-Sasaki compaction function and its significance \label{sec:SS}}

Shibata and Sasaki~\cite{Shibata:1999zs} adopted the spatially conformally flat coordinates
\begin{equation}
 ds^{2}=-\alpha^{2}dt^{2}+\psi^{4}a^{2}(t)[(dr+\beta r dt)^{2}+r^{2}d\Omega^{2}],
\label{eq:CF_coord}
\end{equation}
with the CMC slice in spherical symmetry and 
gave the following expression as an excess in Kodama mass in Eq.~(4.28) in Ref.~\cite{Shibata:1999zs} as
\begin{eqnarray}
 \delta M_{\rm SS}:= 4\pi a^{3}\rho_{0}\int ^{r}_{0}x^{2}dx \delta_{{\rm CMC}} \cdot 
\psi^{6}\left(1+\frac{2x}{\psi}\frac{\partial\psi}{\partial x}\right)
\label{eq:deltaMSS}
\end{eqnarray}
and a compaction function in Eq.~(4.29) in Ref.~\cite{Shibata:1999zs} as
\begin{eqnarray}
 {\cal C}_{\rm SS}(t,r):=\frac{\delta M_{\rm SS}(t,r)}{r\psi^{2}(t,r)a},
\end{eqnarray}
where we put subscript SS (Shibata-Sasaki) to $\delta M$ and ${\cal C}$.
In the long-wavelength solutions, 
${\cal C}_{\rm SS}(t,r)$ becomes time-independent in the limit $\epsilon\to 0$, so that 
\begin{equation}
 {\cal C}_{\rm SS}(t,r)\approx {\cal C}_{\rm SS}(r),
\end{equation}
where the weak equality implies the equality if we neglect higher-order terms in $\epsilon$.
We will show how the Shibata-Sasaki gauge condition fixes the long-wavelength solution including the lapse and shift functions in Sec.~\ref{sec:SS_gauge}. 
We can easily find that ${\cal C}_{\rm SS}$ 
relates to the density perturbation  
averaged inside the radius $r$ at horizon entry $a\Psi^{2}r=H^{-1}$
to the next-to-leading order, which we denote with $\bar{\delta}_{{\rm CMC},H}(r)$, through
\begin{equation}
{\cal C}_{\rm SS}(r)\approx 
\frac{1}{2}\bar{\delta}_{{\rm CMC},H}(r), 
\end{equation}
and this $\bar{\delta}_{{\rm CMC},H}$ has long been used as the indicator 
for the threshold of PBH formation since the pioneering work by Carr~\cite{Carr:1975qj}.
We note that the definition of the compaction function in the long-wavelength solution is physically clearer than that of $\bar{\delta}_{{\rm CMC},H}$ because the latter is not a real value but only obtained by the truncated series expansion.

The profile of the density perturbation, $\delta_{\rm CMC}(t,r)$, is 
typically (but not always) assumed to have the central overdense region 
surrounded by an underdense region. 
Since $\delta_{\rm CMC}(t,r)$ in the long-wavelength solution is given 
in the separable form as seen in Eq.~(\ref{eq:LWL_sol_CMC}), we can 
define $r_{0}$ as the smallest zero of $\delta_{\rm CMC}(t,r)$.
As for the compaction function ${\cal C}_{\rm SS}(r)$, we can usually assume that 
it takes a maximum at some finite radius $r_{\rm max}$. 

Empirically, the maximum of ${\cal C}_{\rm SS}(r)$ gives a good indicator for primordial black hole formation. Shibata and Sasaki~\cite{Shibata:1999zs} 
obtained the threshold value $\simeq 0.4$ 
for radiation $\Gamma=4/3$.
We will concentrate the compaction function, although more recently it has been reported that the maximum value of the volume-averaged 
compaction function is more robust to provide a threshold for black hole formation~\cite{Escriva:2019phb}.

In spherical symmetry, Eq.~(\ref{eq:LWL_sol_CMC}) implies that 
the long-wavelength solution in the CMC slice is given by 
\begin{eqnarray}
\delta_{\rm CMC}&\approx&  f\left(\frac{1}{aH}\right)^{2}, ~
u_{{\rm CMC} j }\approx \frac{2}{3\Gamma(3\Gamma+2)H}\delta_{{\rm CMC},r}\delta^{r}_{j},~~f = -\frac{4}{3}\frac{1}{r^{2}\Psi^{5}}
\left(r^{2}\Psi'\right)'
\label{eq:delta_u_CMC}
\end{eqnarray}
for $\Psi=\Psi(r)$, where the prime denotes the derivative with respect to $r$. 
Using this solution, we can implement the integral in Eq.~(\ref{eq:deltaMSS}) and obtain 
Eq.~(6.33) in Ref.~\cite{Harada:2015yda} or
\begin{equation}
 {\cal C}_{\rm SS}(r)\approx 
\frac{1}{2}\left[1-\left(1+2\frac{d\ln \Psi}{d\ln r}\right)^{2}\right].
\end{equation}
That is, ${\cal C}_{\rm SS}$ is a quadratic function of 
$d\ln \Psi/d\ln r$ and does not include $\Psi''$. 
The threshold is therefore translated to $d\ln \Psi/d\ln r \simeq -0.27$ for radiation, 
noting that ${\cal C}_{\rm SS}$ is a decreasing function 
of $d\ln \Psi/d\ln r$ around $0$.

\section{Shibata-Sasaki compaction function revisited \label{sec:SS_revisited}}

Let us reconsider an excess in mass from the beginning.
In the conformally flat coordinates (\ref{eq:CF_coord}), 
the mass is rigorously given by 
\begin{eqnarray}
 M&=&4\pi \int_{0}^{r}x^{2}dx a^{3}\alpha \psi^{6}T^{t}_{~\mu}K^{\mu} \\
&=&4\pi a^{3}\int^{r}_{0}dx (\psi^{2}x)^{2}
\left\{-[(\rho+p)u^{t}u_{t}+p](\psi^{2}x)'+(\rho+p)u^{t}u_{r}\frac{x}{a}(\psi^{2}a)_{,t}\right\}.
\end{eqnarray}

To define the mass excess, we need to define the background 
mass. For this purpose, let $M_{\rm FF}(t,r)$ denote the mass in the flat 
FLRW spacetime with $\alpha=1$, $\beta=0$ and $\psi=1$.
Then, the mass excess is naturally defined as
\begin{equation}
 \delta M(t,r)=M(t,r)-M_{\rm FF}(t,\psi^{2}(t,r)r), 
\end{equation}
i.e., the difference between masses within the 2-sphere determined by $(t,r)$
in the perturbed spacetime and that of the same area in the background spacetime. 
The mass excess in the CMC slice is then calculated to give
\begin{eqnarray}
&& \delta M_{{\rm CMC}}\approx 4\pi a^{3}\rho_{b}
\int^{r}_{0}dx (\Psi^{2}x)^{2}\left[\delta_{{\rm CMC}} (\Psi^{2}x)'+\frac{2}{3(3\Gamma+2)}\delta_{{\rm CMC}} ' (\Psi^{2}x)\right] \\
&&\hspace{0.5cm}= 4\pi a^{3}\rho_{b}\left[
\frac{3\Gamma}{3\Gamma+2}
\int^{r}_{0}dx (\Psi^{2}x)^{2}(\Psi^{2}x)'
\delta_{{\rm CMC}} +
\frac{2}{3(3\Gamma+2)}\delta_{{\rm CMC}}(t,r) (\Psi^{2}(r)r)^{3}\right],
\end{eqnarray}
where the second equality follows from integration by parts. 
This reduces to
\begin{equation}
 {\cal C}_{\rm CMC}(r)\approx \frac{3\Gamma}{3\Gamma+2}{\cal C}_{\rm SS}(r)+
\frac{1}{3\Gamma+2}f(r)(\Psi^{2}(r)r)^{2}.
\label{eq:CCMC_CSS_f}
\end{equation}
We define
\begin{equation}
 {\cal C}_{\rm CMC}:=\delta M_{{\rm CMC}}/R
\end{equation}
as the legitimate compaction function in the CMC slice.
Since $\delta M_{{\rm SS}}$ is different 
from $\delta M_{\rm CMC}$ 
due to the nonvanishing $ u_{{\rm CMC}j}$, 
$ {\cal C}_{\rm SS}(r)$ is different from ${\cal C}_{\rm CMC}(r)$.

Let us further discuss the relationship between the Shibata-Sasaki and the legitimate CMC compaction functions.
If we assume ${\cal C}_{\rm SS}(r)$ takes a maximum at $r=r_{\rm max}$, then 
the following equation follows: 
\begin{equation}
 \left.\delta_{\rm CMC}(t,r)(\Psi^{2}r)^{3}\right|_{r=r_{\rm max}}=\int_{0}^{r_{\rm max}}dx (\Psi^{2}x)^{2}(\Psi^{2}x)' \delta_{\rm CMC}(t,x).
\end{equation}
This implies
\begin{equation}
 {\cal C}_{\rm CMC}(r_{\rm max})\approx \frac{9\Gamma+2}{3(3\Gamma+2)}{\cal C}_{{\rm SS}}(r_{\rm max}),
\end{equation}
whereas ${\cal C}_{\rm CMC}(r)$ may not take a maximum at $r=r_{\rm max}$ in general. 

Note that ${\cal C}_{\rm SS}(r)$ is unique in the sense that it is written solely in terms of $d\ln \Psi/d\ln r$ at $r$, 
whereas ${\cal C}_{\rm CMC}(r)$ includes a term of $\Psi''(r)$ in addition to a term proportional to ${\cal C}_{\rm SS}(r)$.
We can infer that this is why ${\cal C}_{\rm SS}$ is empirically robust to give a threshold for black hole formation according to the following discussion.
Suppose $f(r)$ or $\Psi''(r)$ vanish except for it has a spike of the order of $\Delta^{-1/2}$ 
at some comoving radius $r=r_{1}>0$ with a small width $\Delta \ll r_{1}$, for which 
the density spike has a large maximum $\propto \Delta ^{-1/2}$ at $r=r_{1}$ with a small width $\Delta$.
This gives a large maximum in ${\cal C}_{\rm CMC}\propto \Delta ^{-1/2}$ at $r=r_{1}$, 
whereas both $\Psi(r)$ and ${\cal C}_{\rm SS}$ are kept very small. 
The small amplitude of curvature perturbation should not induce primordial black hole formation, while 
${\cal C}_{\rm CMC}\propto \Delta ^{-1/2}$ unreasonably has a large maximum in this setting. 

Since the next-to-leading order of the long-wavelength solution depends on the 
time-slicing, the compaction function also does. 
Let us consider the comoving slice, in which the long-wavelength solution 
is given by 
\begin{eqnarray}
\delta_{\rm com}\approx \frac{3\Gamma}{3\Gamma+2}f\left(\frac{1}{aH}\right)^{2},~~
u_{{\rm com} j}=0.
\end{eqnarray}
The mass excess and the legitimate compaction function in the comoving slice 
are calculated to give 
\begin{equation}
 \delta M_{{\rm com}}(t,r)\approx \frac{3\Gamma}{3\Gamma+2}\delta M_{\rm SS}(t,r)
\end{equation}
and 
\begin{equation}
 {\cal C}_{\rm com}(r):=\frac{\delta M_{\rm com}}{R}\approx \frac{3\Gamma}{3\Gamma+2}{\cal C}_{\rm SS}(r), 
\end{equation}
respectively. Thus, we can interpret 
${\cal C}_{\rm SS}(r)$ as $({3\Gamma+2})/({3\Gamma})$ times 
${\cal C}_{\rm com}(r)$. Therefore, the maximum value of 
${\cal C}_{\rm com}(r)$ is given by ${\cal C}_{\rm com}(r_{\rm max})$.
The threshold value for the maximum value of ${\cal C}_{\rm com}(r)$ is 
$0.4\times 2/3\simeq 0.27$ for radiation $\Gamma=4/3$.

Similarly, we can calculate the legitimate compaction functions 
in different time slices including the geodesic slice denoted by G 
and the uniform-density slice denoted by UD. 
Although we relegate detailed calculations to Appendix~\ref{sec:G_UD_slices},
we can find that except for ${\cal C}_{\rm com}(r)$, the legitimate compaction functions in these slices 
contain not only $d\ln \Psi/d\ln r$ at $r$ but also $\Psi''(r)$ and therefore have no 
direct relation  to ${\cal C}_{\rm SS}(r)$. 
If we choose $r=r_{0}$, which is a zero of $f$, we find that 
all of 
${\cal C}_{{\rm com}}(r_{0})$, ${\cal C}_{{\rm CMC}}(r_{0})$, 
${\cal C}_{G}(r_{0})$ and ${\cal C}_{{\rm UD}}(r_{0})$ 
coincide with each other and are equal to 
$[{3\Gamma}/({3\Gamma+2})]{\cal C}_{\rm SS}(r_{0})$.
Thus, the legitimate compaction functions at $r=r_{0}$ are gauge-independent.
If we take $r=r_{\rm max}$ instead, where ${\cal C}_{\rm SS}(r)$ takes a maximum, 
the values of the compaction functions in different slices there 
can be expressed by ${\cal C}_{\rm SS}(r_{\rm max})$ times 
different constant factors depending on $\Gamma$.

The definition of the mass excess needs the definition of the background 
mass in the flat FLRW spacetime. 
We think that the choice of the sphere with the same area is most natural but 
alternative natural choice would be the {\it comoving} sphere with as close
area as possible. We will relegate the discussion on this alternative choice to 
Appendix~\ref{sec:alternative_background},
in which it is shown that the Shibata-Sasaki compaction function cannot be 
the ratio of thus defined mass excess to the areal radius, either.

\section{Long-wavelength solutions in the Shibata-Sasaki gauge condition\label{sec:SS_gauge}}

The full set of long-wavelength solutions to the next-to-leading order are obtained in Ref.~\cite{Harada:2015yda} 
but only in the normal coordinates for the threading condition, 
while Shibata and Sasaki adopted 
the CMC slice and the conformally flat coordinates.
Generally speaking, this makes 
it very difficult to compare the result of Shibata and Sasaki with most of the other works 
mainly based on the gauge condition of the comoving slice and comoving thread.
So, we here construct the 
long-wavelength solutions in the gauge conditions Shibata and Sasaki adopted.

In the CMC condition, $\delta$ and $u_{\mu}$ are given by 
Eq.~(\ref{eq:delta_u_CMC}). As for $\chi$, 
Eq.~(4.39) in Ref.~\cite{Harada:2015yda} with $\kappa=0$ gives
\begin{equation}
 \chi\approx -\frac{3\Gamma-2}{3\Gamma}f\frac{1}{(aH)^{2}}.
\label{eq:chi}
\end{equation}
On the other hand, $\beta^{r}=r\beta$ is determined by the 
conformally flat coordinate condition. In fact, 
Eqs.~(4.19), (4.28) and (4.31) in Ref.~\cite{Harada:2015yda} with $h_{ij}=0$ imply
\[
\tilde{A}_{22}\approx \frac{2}{3\Gamma+2}\left(\frac{1}{\Psi^{4}}{\cal C}_{\mathrm{SS}}-\frac{1}{2}r^{2}f \right)\frac{1}{a^{2}H}
\]
and
\begin{equation}
 \beta'\approx -\frac{6}{3\Gamma+2}\frac{1}{r^{3}}
\left(\frac{1}{\Psi^{4}}{\cal C}_{\rm SS}-\frac{1}{2}r^{2}f\right)\frac{1}{a^{2}H}, 
\label{eq:SS_beta'}
\end{equation}
where $\tilde{A}_{ij}$ is defined in Ref.~\cite{Harada:2015yda}.
Integrating the above gives
\begin{equation}
 \beta\approx \left\{-\frac{6}{3\Gamma+2}
\int_{\infty}^{r}d\tilde{r}\frac{1}{\tilde{r}^{3}}
\left[\frac{1}{\Psi^{4}(\tilde{r})}{\cal C}_{\rm SS}(\tilde{r})-\frac{1}{2}\tilde{r}^{2}f(\tilde{r})\right]
+\tilde{\beta}_{\infty}\right\}\frac{1}{a^{2}H}
=:\tilde{\beta}(r)\frac{1}{a^{2}H},
\end{equation}
where $\tilde{\beta}_{\infty}$ is a constant of integration and regarded as 
a remaining gauge freedom.
The above fixes the lapse and shift to the next-to-leading order.

Now we should obtain $\xi$. 
Since Eq.~(4.16) in Ref.~\cite{Harada:2015yda} with $\kappa=0$ and Eq.~(\ref{eq:chi}) 
requires $\xi\propto t^{2-({4}/{3\Gamma})}$, where the time-independent 
part of $\xi$ is absorbed into $\Psi(r)$, 
we find 
\begin{equation}
 \dot{\xi}\approx (3\Gamma-2)H\xi, 
\label{eq:dotxi}
\end{equation}
so that Eq.~(4.16) in Ref.~\cite{Harada:2015yda} yields
\begin{equation}
 -H\chi +2(3\Gamma-2)H\xi -\beta \left(1+\frac{2r\Psi'}{\Psi}\right)-\frac{1}{3}r\beta'\approx 0.
\label{eq:HYNK4.16'}
\end{equation}
Therefore, we obtain
\begin{eqnarray}
 \xi&\approx& \frac{1}{2(3\Gamma-2)}\left\{\chi+\frac{1}{H}\left[\frac{1}{3}r\beta'+
\beta\left(1+\frac{2r\Psi'}{\Psi}\right)\right]\right\} \nonumber \\
 &\approx &
\frac{1}{2(3\Gamma-2)}
\left\{
-\frac{2}{3\Gamma+2}\frac{{\cal C}_{\rm SS}}{r^{2}\Psi^{4}}
-\frac{9\Gamma^{2}-3\Gamma-4}{3\Gamma(3\Gamma+2)}f
+\left(1+\frac{2r\Psi'}{\Psi}\right)\tilde{\beta}(r) \right\}
\frac{1}{a^{2}H^{2}} \nonumber \\
&=:&\tilde{\xi}(r)\frac{1}{a^{2}H^{2}}.
\end{eqnarray}

Now that we have the solution for all the perturbation quantities, 
$\chi$, $\beta$, $\delta$, $u_{r}$, $\kappa$ and $\xi$, 
let us now check consistency in terms of the compaction function.
As we have seen, the compaction function is given not only in terms of the spatial integral
but also in terms of the metric functions at the surface of the volume.
To see this, we begin with the definition (\ref{eq:M_MS}), where $R(t,r)=\psi^{2}(t,r)a(t)r$, as seen in the metric (\ref{eq:CF_coord}). The  result is
\begin{eqnarray}
 \frac{2M}{R}&=&1+
\frac{1}{\alpha^{2}}\left(H+2\frac{\dot{\psi}}{\psi}\right)^{2}(\psi^{2}ra)^{2}
-2\frac{\beta}{\alpha^{2}}\left(H+2\frac{\dot{\psi}}{\psi}\right)\frac{(\psi^{2}r)'}{\psi^{2}}(\psi^{2}ra)^{2} \nonumber \\
&& -\frac{(\psi^{2}r)'^{2}}{\psi^{4}}+\frac{\beta^{2}}{\alpha^{2}}\frac{(\psi^{2}r)'^{2}}{\psi^{4}}(\psi^{2}ra)^{2}.
\end{eqnarray}
For the long-wavelength solution, subtracting the contribution from the background mass, 
$ {2M_{\rm FF}}/{R}=H^{2}R^{2}$, we find 
 \begin{equation}
 \frac{2\delta M}{R}\approx 1-\left(1+\frac{2r\Psi'}{\Psi}\right)^{2}  
+2\left[2\dot{\xi}-H\chi-\beta \left(1+\frac{2r\Psi'}{\Psi}\right)\right](\Psi^{2}r )^{2} (a^{2}H)
\label{eq:2deltaM/R}
\end{equation}
where we have kept only the terms which remain in the limit $t\to 0$.
Using Eqs.~(\ref{eq:SS_beta'}), (\ref{eq:dotxi}) and (\ref{eq:HYNK4.16'}), 
we find that Eq.~(\ref{eq:2deltaM/R}) reduces to 
\begin{equation}
{\cal C}_{\rm CMC}(r) \approx \frac{3\Gamma}{3\Gamma+2}
{\cal C}_{\rm SS}(r)
+\frac{1}{3\Gamma+2}f(r) (\Psi^{2}(r)r)^{2}.
\end{equation}
This coincides with the expression (\ref{eq:CCMC_CSS_f}), 
which is obtained using the spatial integral.
To show this consistency, specifying $\beta'$ using 
the threading condition is crucial.

\section{Summary \label{sec:summary}}

We have shown that in spite of the initial intention, 
the Shibata-Sasaki compaction function in the long-wavelength solution 
does not coincide with the legitimate compaction function $\delta M/R$ in terms of the 
standard definition of a quasi-local mass. 
The misinterpretation did not only remain in Ref.~\cite{Shibata:1999zs} but 
also propagated to Ref.~\cite{Harada:2015yda}.  
Nevertheless, we should stress that the Shibata-Sasaki compaction function remains particularly
useful to give a robust threshold for primordial black hole formation. This is due to its property 
that it is directly related to the curvature perturbation and 
its first spatial derivative but not its second.
This property is shared by the legitimate compaction function only in the comoving slice, which would give another physical interpretation of the Shibata-Sasaki compaction function.

\acknowledgements

The authors are grateful to I.~Musco, K.~I.~Nakao and M.~Sasaki for helpful comments.
This work was partially supported by JSPS KAKENHI Grant No. 
JP19H01895 (TH, CY, YK), No. JP19K03876(TH), No. JP20H05850 (CY), No. JP20H05853 (TH, CY, YK), No. JP21K20367 (YK).
\appendix

\section{Legitimate compaction functions in different slices
\label{sec:G_UD_slices}}

In the long-wavelength solutions in the geodesic and uniform-density slices, 
$\delta$ and $u_{j}$ are respectively given by
Eqs. (4.74) and (4.77) in Ref.~\cite{Harada:2015yda} or 
\begin{eqnarray}
 \delta _{G} \approx \frac{3\Gamma}{9\Gamma-4}f\frac{1}{(aH)^{2}}, \quad 
u_{G j} \approx -\frac{6(\Gamma-1)}{(9\Gamma-4)(3\Gamma+2)}f_{,j}\frac{1}{H(aH)^{2}}, 
\end{eqnarray}
and 
\begin{eqnarray}
\delta_{{\rm UD}} \approx 0, \quad u_{\mathrm{UD}j} \approx -\frac{1}{3\Gamma+2}f_{,j}\frac{1}{H(aH)^{2}}, 
\end{eqnarray}
where $f$ is defined in Eq.~(\ref{eq:f}).
Further in spherical symmetry, mass excesses are calculated to 
\begin{eqnarray}
 \delta M_{G}(t,r)\approx \frac{3\Gamma}{3\Gamma+2}\frac{4\pi a^{3}\rho_{b}}{(aH)^{2}}\left[\int^{r}_{0}dx (\Psi^{2}x)^{2}(\Psi^{2}x)' f
-\left.\frac{2(\Gamma-1)}{9\Gamma-4}(\Psi^{2}x)^{3}f\right|_{x=r}\right], 
\end{eqnarray}
and 
\begin{eqnarray}
 \delta M_{\rm UD}(t,r) \approx \frac{3\Gamma}{3\Gamma+2}\frac{4\pi a^{3}\rho_{b}}{(aH)^{2}}\left[\int^{r}_{0}dx (\Psi^{2}x)^{2}(\Psi^{2}x)' f 
-\left.\frac{1}{3}(\Psi^{2}x)^{3}f\right|_{x=r}\right], 
\end{eqnarray}
respectively. It is curious that all of the legitimate compaction functions agree 
if the upper bound $r$ of the integral is chosen to $r_{0}$, 
where $\delta _{{\rm CMC}}(t,r_{0})=f(r_{0})=0$, 
that is,  
\begin{equation}
 \delta M_{\rm com}(t,r_{0})
\approx \delta M_{\rm CMC}(t,r_{0})
\approx \delta M_{G}(t,r_{0})
\approx \delta M_{\rm UD}(t,r_{0})
\approx \frac{3\Gamma}{3\Gamma+2}M_{\rm SS}(r_{0}),
\end{equation}
and thus we have
\begin{equation}
 {\cal C}_{\rm com}(r_{0}) \approx {\cal C}_{\rm CMC}(r_{0})
\approx {\cal C}_{G}(r_{0})
\approx {\cal C}_{\rm UD}(r_{0}) \approx 
\frac{3\Gamma}{3\Gamma+2}{\cal C}_{\rm SS}(r_{0}).
\end{equation}
However, for general $r$, there is no such simple relation between them.

\section{Alternative choice of the mass excess \label{sec:alternative_background}}

An alternative reasonable choice of the background mass 
would be that contained within the {\it comoving} sphere of almost the same 
area, in which the time evolution of the background mass is trivial. 
This promotes us to alternatively choose $M_{\rm FF}(t,\Psi^{2}(r)r)$ 
rather than $M_{\rm FF}(t,\psi^{2}(t,r)r)$ for the background mass. 
This gives the definition of the alternative mass excess and compaction function
in the CMC slice:
\begin{eqnarray}
 \Delta M(t,r)&:=& M(t,r)-M_{\rm FF}(t,\Psi^{2}(r)r), \\
 \mathscr{C}_{\rm CMC}(t,r)&:=&\frac{\Delta M (t,r)}{R(t,r)}\approx \mathscr{C}_{\rm CMC}(r).
\end{eqnarray}
In this section, we will always restrict ourselves to the CMC slice, so we will omit the suffix `CMC'.
Since the relationship between $\Delta M$ and $\delta M$ is given by 
\begin{equation}
\Delta M=\delta M+8\pi a^{3}\rho_{b}(\Psi^{2}r)^{3}\xi,
\end{equation}
the compaction function $\mathscr{C}:=\Delta M/R$ can be written as
\begin{equation}
 \mathscr{C}(r)-{\cal C}(r)=3(\Psi^{2}(r)r)^{2}(aH)^{2}\xi.
\end{equation}
Moreover, using Eq.~(\ref{eq:CCMC_CSS_f}), we obtain
\begin{equation}
 \mathscr{C}(r)-{\cal C}_{\rm SS}(r)=
-\frac{2}{3\Gamma+2}{\cal C}_{SS}(r)+\frac{1}{3\Gamma+2}f(r)(\Psi^{2}(r)r)^{2}+3\xi
(aH)^{2}(\Psi^{2}(r)r)^{2}.
\label{eq:C-C_SS}
\end{equation}

To show that neither $\mathscr{C}(r)= {\cal C}(r)$ 
nor $\mathscr{C}(r)= {\cal C}_{SS}(r)$ holds in general,  
let us consider linear perturbation from the flat FLRW solution by putting 
$\Psi(r)=1+h(r)$. Linearising ${\cal C}_{\rm SS}(r)$ and $f(r)$ with respect to $h$, we find 
\begin{eqnarray}
 {\cal C}_{\rm SS}(r)&\approx& -2rh', \\
 f(r)&\approx& -\frac{4}{3}\frac{1}{r^{2}}(r^{2}h')'.
\end{eqnarray}
Choosing the function $h(r)$ as
\begin{equation}
h(r)=
\begin{cases}
-\frac{1}{8}f_{0}(r^{2}-3r_{0}^{2}) & (0<r<r_{0}) \\
\frac{1}{4}f_{0}r_{0}^{3}\displaystyle\frac{1}{r} & (r_{0}\le r),
\end{cases}
\end{equation}
we find 
\begin{eqnarray}
 {\cal C}_{\rm SS}(r)&\approx & 
\begin{cases}
 \frac{1}{2}f_{0}r^{2} & (0<r<r_{0}) \\
 \frac{1}{2}f_{0}r_{0}^{3}\displaystyle\frac{1}{r} & (r_{0}\le r)
\end{cases} \\
f(r) &\approx & 
\begin{cases}
f_{0} & (0<r<r_{0}) \\
0 & (r_{0}\le r)
\end{cases} \\
r\tilde{\beta}'(r)&\approx &
\begin{cases}
 0 & (0<r<r_{0}) \\
 -\frac{3}{3\Gamma+2}f_{0}\left(\frac{r_{0}}{r}\right)^{3}& (r_{0}\le r)
\end{cases}
\label{eq:rtildebeta'}\\
 \tilde{\beta}(r)&\approx &
\begin{cases}
 \frac{1}{3\Gamma+2}f_{0}+\tilde{\beta}_{\infty}& (0<r<r_{0}) \\
 \frac{1}{3\Gamma+2}f_{0}\left(\frac{r_{0}}{r}\right)^{3}+\tilde{\beta}_{\infty}& (r_{0}\le r)
\end{cases} \\
 \tilde{\xi}(r)&\approx& 
\begin{cases}
\frac{1}{2(3\Gamma-2)}\left[-\frac{9\Gamma^{2}-3\Gamma-4}{3\Gamma(3\Gamma+2)}f_{0}
+\tilde{\beta}_{\infty} \right]& (0<r<r_{0}) \\
\frac{1}{2(3\Gamma-2)}\tilde{\beta}_{\infty}&(r_{0}\le r),
\end{cases}
\end{eqnarray}
where $\tilde{\beta}_{\infty}$ is assumed to be of the 
order of $h$.
Since $\tilde{\xi}(r)$ depends on the gauge constant $\tilde{\beta}_{\infty}$, 
$\mathscr{C}-{\cal C}=3(\Psi^{2}(r)r)^{2}\tilde{\xi}$ also does. 
However, even if we choose 
$\tilde{\beta}_{\infty}=0$ so that $\mathscr{C}={\cal C}$ 
for $r_{0}\le  r $, we find $\mathscr{C}\ne {\cal C}$ in general for $0<r<r_{0}$.
Also from Eq.~(\ref{eq:C-C_SS}), we find 
\begin{equation}
\mathscr{C}(r)- {\cal C}_{\rm SS}(r)\approx 
\begin{cases}
\frac{1}{2(3\Gamma-2)}\left[-\frac{9\Gamma^{2}-3\Gamma-4}
{\Gamma(3\Gamma+2)}f_{0}+3\tilde{\beta}_{\infty}
\right] r^{2} & (0<r<r_{0}) \\
-\frac{1}{3\Gamma+2}f_{0}r_{0}^{3}\frac{1}{r}+\frac{3}{2(3\Gamma-2)}\tilde{\beta}_{\infty}r^{2} & (r_{0}\le r)
\end{cases}
\end{equation}
Thus, we can also conclude $\mathscr{C}\ne {\cal C}_{\rm SS}$ in general.

\end{document}